# Ordering/displacive ferrielectricity in 2D $CuInP_2S_6$

Yulian Vysochanskii[a*], Vitalii Liubachko[a], Konstantin Glukhov[a], Ruslan Yevych[a], Anton Kohutych[a], Viacheslav Hryts[a], Andrius Dziaugys[b], Juras Banys[b]

[a]*Institute for Solid State Physics and Chemistry, Uzhhorod University, Pidgirna Str. 46, Uzhhorod, 88000, Ukraine;* [b]*Faculty of Physics, Vilnius University, Sauletekio 9, Vilnius, 10222, Lithuania*

*corresponding author: vysochanskii@gmail.com

Provide short biographical notes on all contributors here if the journal requires them.

# Ordering/displacive ferrielectricity in 2D $CuInP_2S_6$

Using DFT-based molecular dynamics simulation of $Cu^+$ cations flipping and $In^{3+}$ cations displacive dynamics, we clarify the dipole ordering of $CuInP_2S_6$ ferrielectrics through the second order Jahn-Teller effect which determines the double-well potential for copper cations as well as three-well potential for indium cations inside the structural layers. The temperature dependence of the spatial distribution of $Cu^+$ and $In^{3+}$ cations is also described within the quantum anharmonic oscillators model. $In^{3+}$ cations play a decisive role in the character of the polar ordering: at ambient or for positive pressures, the first order transition occurs; with the rise of negative compression, the ferrielectric transition evolves to the second order. The peculiarities of spontaneous polarization switching are related to the contributions of copper and indium cationic sublattices.

## Introduction

Ferrielectric polarization in van der Waals (vdW) $MM'P_2S(Se)_6$ (M = Cu, M' = In) crystals is determined by opposite shifts of $Cu^+$ and $In^{3+}$ cations out of the middle of the structural layers that are built by $(P_2S_6)^{4-}$ anionic structural groups. Such structural ordering below $T_c$ can be represented as the freezing of the hopping motion of $Cu^+$ cations in double-well potential and the displacement of $In^{3+}$ cations when the temperature decreases [1]. The second order Jahn-Teller effect destabilizes the $Cu^+$ cations in the quasi-octahedral positions in the middle of the structural layers [2]. In addition, the interlayer interaction promotes the thermal throw of $Cu^+$ cations into the vdW gap at high temperatures [1].

Dielectric studies of $CuInP_2S_6$ crystal illustrate the presence of the first order phase transition near $T_c \sim 315$ K with the critical slowdown of the relaxational soft mode, which is appropriate for the order-disorder type phase transition [3]. Using DFT-based calculations for selenide analog $CuInP_2Se_6$, it was found, that the contribution to the total spontaneous polarization by polar modes of $Cu^+$ and $In^{3+}$ cations is dominant [4]. But the nonlinear coupling of these modes with the fully symmetric breathing mode of $(P_2Se_6)^{4-}$ anions provides energetic stabilization of the ferrielectric phase in the three-well-shaped total energy surface [4]. A similar three-well local potential for the spontaneous polarization fluctuations arises in $Sn_2P_2S_6$ ferroelectric 3D crystal due to the $Sn^{2+}$ cations $5s^2$ stereochemically active electron lone pair [5]. In the case of 2D $CuInP_2S(Se)_6$ crystals, the three-well local potential for $In^{3+}$ cations is also appropriate because of the stereochemical activity, that follows from the $sp^2$ hybridization of indium and sulfur valence orbitals [6]. This circumstance was overlooked earlier [2], only a "geometrical" origin of the $In^{3+}$ cations displacement was supposed.

Using *ab initio* molecular dynamics simulations, quantum anharmonic oscillator model analysis, and mean-field analysis we clarify the different roles of $Cu^+$ ordering and $In^{3+}$ displacive dynamics in the spontaneous polarization of $CuInP_2S_6$ ferrielectric crystals. The contributions of $Cu^+$ and $In^{3+}$ sublattices into ferrielectric multipolar states are related to the topology of the $CuInP_2S_6$ temperature-pressure phase diagram. The transformation of spontaneous polarization switching across the temperature-pressure phase diagram is discussed in comparison with experimental data about the evolution of the ferrielectric ordering at chemical substitution in $Ag_{0.1}Cu_{0.9}InP_2S_6$ and $CuInP_2(Se_{0.02}S_{0.98})_6$ mixed crystals.

## *Ab initio* molecular dynamics simulations

The CASTEP software package was used for molecular dynamics (MD) simulation of the temperature behavior of the $CuInP_2S_6$ ferrielectric. A supercell consisting of 2x2x1 primitive unit cells of $CuInP_2S_6$ crystal has been considered (Figure 1). Initially, the $CuInP_2S_6$ unit cell was optimized at $T = 0$ K. Geometrical optimization was performed by minimization of atomic forces and cell total energy, calculated by means of DFT within generalized gradient approximation (GGA) with dispersion-corrected functional for including van der Waals interaction. Ultrasoft pseudopotentials [7] were used throughout the calculation. Electronic configurations for ions present in the system are the following: Cu – $3d^{10}4s^{1}$; In – $3d^{10}4s^{2}4p^{1}$; P – $3s^{2}3p^{3}$; S – $3s^{2}3p^{4}$. MD is based on the velocity Verlet algorithm for the integration of the equation of motion. Constant pressure and temperature ensemble (NPT) with Nose thermostat, and Andersen method of pressure control (barostat) were used to consequent stepwise "heating" of the system during the simulation from 77 K up to 390 K. Timestep after some test runs was set to 1 fs, and the total number of steps (time) for each temperature was equal to 2000 (2 ps). An ionic motion was unrestricted, and symmetry conditions were not imposed. In Figure 1 the displacements of ions from the midplane of the structural layers are averaged over simulation time for each temperature. The probability is proportional to the time spent in the position and the depth of the potential at this point.

For each temperature, the spatial distribution of ions within structural layer is presented by the corresponding curves. Accordingly, the higher and the narrower the curve is the more the corresponding ion is localized. As it follows from Figure 1, at low temperatures, the ions within the layer are well localized at their sites, with the exception of Cu ions, which perform low-amplitude movements limited by the planes of S- and P-ions. At the same time, the position of In ions does not coincide with the center of the layer and is shifted in the direction opposite to the position of Cu ions. At higher temperatures, Cu ions become more liable and demonstrate penetration to the inner (between phosphor planes) and outer (beyond the sulfur plane in the vdW gap) space.

At the highest temperatures, some of Cu ions can be located near both the lower and upper boundaries of the structural layer, which corresponds to the tendency of the transition to a centrosymmetric state.

MD modeling shows that the temperature dependence of the spatial distribution of $Cu^+$ cations can be described by the double-well local potential (Figure 2) which is determined by their “up” and “down” stable positions at low temperatures. An additional thin barrier inside the potential well reflects the separation of the Cu1 positions inside the structural layer and Cu3 positions located in the vdW gap. A three-well potential with metastable side wells is suitable for $In^{3+}$ cations.

## Analysis within the quantum anharmonic oscillator model

We have applied the quantum anharmonic oscillator (QAO) model [8] which considers phonon-like bosonic excitations. These excitations can be regarded as the incorporation of electron-phonon interactions. In the QAO model, the $CuInP_2S_6$ crystal is presented (see Figure 2(a)) as the chain of quantum anharmonic oscillators associated with $Cu^+$ and $In^{3+}$ cations in two parallel structural layers. The Hamiltonian of such a system is:

$$H = \sum_i \left( T(x_i) + V(x_i) + J \langle x \rangle x_i \right), \tag{1}$$

where $T(x_i)$ and $V(x_i)$ are operators of kinetic and potential energy, respectively, $\langle x \rangle$ is the average value of the displacement coordinate $x_i$, and $J$ is a coupling constant. The masses of oscillators were equal to copper and indium atomic masses. The last term in Hamiltonian reflects the mean-field approach by taking into account the relation

$$\sum_{ij} J_{ij} x_i x_j \approx \sum_i J \langle x \rangle x_i . \tag{2}$$

The total system consists of two subsystems for Cu and In ions, coupling constants $J_{\text{Cu-Cu}}$ and $J_{\text{In-In}}$ within subsystems and $J_{\text{Cu-In}}$ between subsystems: $H_{\text{total}} = H_{\text{Cu}} + H_{\text{In}}$, here

$$H_{\text{Cu}} = \sum_i \left( T_{\text{Cu}}(x_i) + V_{\text{Cu}}(x_i) + (J_{\text{Cu-Cu}} \langle x \rangle + J_{\text{Cu-In}} \langle y \rangle) x_i \right), \tag{3}$$

and

$$H_{\text{In}} = \sum_i \left( T_{\text{In}}(y_i) + V_{\text{In}}(y_i) + (J_{\text{Cu-In}} \langle x \rangle + J_{\text{In-In}} \langle y \rangle) y_i \right). \tag{4}$$

In the applied model, the effective particle is an oscillator under the influence of a symmetry-breaking field which is calculated self-consistently. Solving Schrödinger equation with Hamiltonian $H_{\text{total}}$, one obtains a set of eigen energies of levels and its wave functions which are used for self-consistent calculation of average expectation value. It should be noted that temperature is included in the Hamiltonian $H_{\text{total}}$ indirectly through Boltzmann distribution for the occupation of the levels. From the definition of $\langle x \rangle$, it follows $\langle x \rangle = \sum_n p_n x_n$, where $p_n$ is the occupation number for the n$^{\text{th}}$ level, $x_n = \int \Psi_n^* x \Psi_n dx / \int \Psi_n^* \Psi_n dx$ is an average value of displacement for the n$^{\text{th}}$ level.

In the paraelectric phase for average displacement $\langle x \rangle$, which is proportional to the order parameter, one can obtain $\langle x \rangle = 0$, and below phase transition temperature $\langle x \rangle \neq 0$. To study the influence of an external field $E$ on the system under consideration, the term proportional to $Ex_i$ should be included into system Hamiltonian.

The shape of the double-well local potential for $Cu^+$ cations was presented by the even power function of displacement $x$, like

$$U = -\text{a}x^8 + \text{b}x^{10}, \tag{5}$$

with the set of coefficients that permit modeling the temperature dependence of the disordering of cooper cations, with a spatial distribution that matches with experimentally observed one by electron diffraction [1] (Figure 3). Gaussian-like terms were also added to the potential (Equation 5) to model the potential barriers associated with the movement of Cu ions from the interlayer space into the vdW gap (Figure 2c). The three-well local potential with metastable side wells was used for the simulation of displacive dynamics of $In^{3+}$ cations. Calculated displacements at 0 K were normalized by a deviation of 1.55 Å away from the middle of the structural layer for $Cu^{+}$ cations, and by a deviation of 0.24 Å for $In^{3+}$ cations, according to the structural data [1]. The contributions of cations to the spontaneous polarization (order parameter) were found by multiplying the calculated displacement by nominal ionic charges +1 for copper and +3 for indium.

It is seen (Figure 4) that for selected parameters of the QAO model, the calculated temperature transformation of the spatial distribution of $Cu^{+}$ cations for stable solution **1** agrees with the experimentally observed (Figure 3) symmetrical distribution relative to the layer midplane in the paraelectric phase. The $In^{3+}$ cations show a tendency to shift to center of the layer with increasing temperature (Figure 4). For the metastable solution **2** at $T_c$ = 315 K, the ions positions coincide with those found for solution **1,** and the temperature behavior of the spatial distribution of $Cu^{+}$ cations is the same as shown above for stable solution **1**. For the solution **2,** the $In^{3+}$ cations below $T_c$ are in the potential with two wells, which are separated by a small barrier (Figure 5). At low temperatures, the spatial distribution of $In^{3+}$ cations is very smeared and shows complex transformations when heated to $T_c$.

For metastable solution **3** (Figure 3c), the calculated order parameter continuously decreases at heating, demonstrating the possibility of the second order phase transition at $T_0 = 275$ K. In this case of solution **3,** the $In^{3+}$ cations are slightly shifted from the center of the structural layer and demonstrate a smeared spatial distribution above 200 K.

The calculations performed in the QAO model for the contributions of $Cu^+$ and $In^{3+}$ cations to the spontaneous polarization $P$ switching by the external electric field $E$ for $CuInP_2S_6$ crystal at 200 K are illustrated in Figure 6. The three-loops shape of the $P$-$E$ hysteresis follows from the coupling between $Cu^+$ and $In^{3+}$ cationic sublattices and switching process depending on the mixing of stable solution **1** with metastable **2** and **3** solutions. When the electric field oriented along the induced $Cu^+$ dipoles increases, both the stable **1** and metastable **2** solutions related to the ferrielectric first order transition coincide. At this, the displacement of $In^{3+}$ cations related to the second order transition (metastable solution **3**) approaches zero, and at a higher electric field takes the opposite orientation, i.e., a transition from the ferrielectric to ferroelectric state occurs.

**Discussion of results**

Let us compare the atomistic consideration of spontaneous polarization reversal for $CuInP_2S_6$ ferrielectrics with the analysis in the mean-field approach [9]. At negative hydrostatic pressures $\sigma = \sigma_1 + \sigma_2 + \sigma_3$, the temperature of the first order paraelectric to ferrielectric transition decreases and after reaching of the critical end point (CEP) at $\sigma = -33$ MPa the phase transitions line splits: the second order paraelectric-to-ferrielectric transitions line and the first order isostructural transitions line appear. Isostructural transitions occur between ferrielectric states FI1 and FI2. Two ferrielectric states differ in the amplitude of the spontaneous polarization: in the FI2 region between the continuous paraelectric-ferrielectric phase transition and the first order ferrielectric-ferrielectric transition the spontaneous polarization is smaller than in the FI1 region below the ferrielectric-ferrielectric transition. The line of the isostructural first order transitions inside the ferrielectric phase terminates at a bicritical end point (BEP) near -113 MPa [9].

The ferrielectric FI1-paraelectric first order transition temperature increases and the FI1 state region expands at compressive hydrostatic stress σ, and this is mainly expressed at uniaxial stress $\sigma_3$ (normally to the structural layers) [9]. This peculiarity is related to the promotion of thermal throw into the vdW gap of $Cu^+$ cations by $\sigma_3$ stress which determines the negative sign of electrostriction in van der Waals $CuInP_2S_6$ crystals, and the growth of its spontaneous polarization at compression [10]. Induced biaxial negative stress $\sigma_2 + \sigma_3$ stretching in the plane of $CuInP_2S_6$ structural layer increases the second order ferrielectric FI2-paraelectric transition temperature and widens the range of the FI2 ferrielectric state with small spontaneous polarization [9].

The topology of the temperature-pressure phase diagram of $CuInP_2S_6$ and the features of the shape of the *P-E* hysteresis loops are related to the interplay between the polar and antipolar ordering of the crystal lattice [9,11]. More specifically, the peculiarities of spontaneous polarization reversal follow from the coupling between the sublattices of $Cu^+$ and $In^{3+}$ cations which play different leading roles. While heating and at normal to the structural layer stress, the ordering dynamics of $Cu^+$ cations is accompanied by their thermal throw into the vdW gap which determines the negative electrostriction [10] and ferroionic properties [3] of $CuInP_2S_6$ crystals. The displacive anharmonic dynamics of $In^{3+}$ cations occurs at the transformation from the first order into the second order of the paraelectric-ferrielectric transition at in-plane biaxial tensile stress and in the regions of stability of the ferrielectric states FI1 and FI2.

Qualitative change in the structural evolution of $CuInP_2S_6$ at cooling below 200 K (change from positive to negative value of thermal expansion coefficient [12], peculiar temperature behavior of lattice vibrations [12,13]) and coexistence of the ferrielectric phase with a dipole glass state at low temperatures [14] correlate with the differences in the anharmonic displacive dynamics of $In^{3+}$ cations in the temperature range below 200 K and in the higher temperature interval shown in Figure 5. This demonstrates the importance of the anharmonic displacive dynamics of $In^{3+}$ cations for the physical properties of the family of $CuInP_2S_6$ ferrielectric crystals.

As it was shown earlier [15], the splitting of the heat capacity anomaly in $Ag_{0.1}Cu_{0.9}InP_2S_6$ mixed crystal can be interpreted as a transformation of the ferrielectric first order phase transition, that is observed for $CuInP_2S_6$, into the sequent paraelectric-ferrielectric second order phase transition and the isostructural first order transition between two ferrielectric states with a significant change of the spontaneous polarization at cooling below 290 K. The shape of the three-well local potential for $In^{3+}$ cations inside sulfur octahedron at Cu by Ag substitution changes due to internal chemical pressure that can be compared with the influence of external mechanical stretching along the structural layers of $CuInP_2S_6$ crystal.

Partial substitution of sulfur by selenium in the anionic sublattice of $CuInP_2(Se_xS_{1-x})_6$ mixed crystals can also induce internal negative chemical pressure that transforms the shape of the three-well local potential for $In^{3+}$ cations inside the octahedron of chalcogen atoms. The rise of selenium concentration rapidly lowers the temperature $T_c$ of the paraelectric-to-ferrielectric phase transition which is similar to the stretching influence - in $CuInP_2(Se_{0.02}S_{0.98})_6$ crystal the ferrielectric state appears below $T_c \sim 290$ K [16].

The thermal throw of $Cu^+$ cations from quasi-trigonal positions (in the structural layer) into quasi-tetrahedral positions (in the vdW gap) is related to the ionic conductivity and also produces dipole defects of the crystal lattice. These defects can pin the spontaneous polarization below 250 K, where the thermal energy is too small to activate the ionic conductivity [17]. At low temperatures, for example at 200 K (Figure 7a), only displacements of $In^{3+}$ cations are involved in the observed polarization reversal at small electric fields. At the temperature increasing from 250 K, the $Cu^+$ cations depinning promotes their involvement in the switching process, which appears at the opening of *P-E* loops (Figure 7a).

In the case of $CuInP_2(Se_{0.02}S_{0.98})_6$ crystal, the observed *P-E* loop at 249.3 K (Figure 7b) [16] may be similar to the calculated triple loop [11]. Here, at small electric fields, the displacement of $In^{3+}$ cations mostly contributes to the lattice polarization. At strong electric fields, the $Cu^+$ sublattice is also involved in the switching. For $Ag_{0.1}Cu_{0.9}InP_2S_6$ crystal, the observed piezoelectric signal at 260 K (Figure 7c) [18] clearly shows the polarization switching process in the FI2 state with a small value of the spontaneous polarization, that can be related to the displacement of $In^{3+}$ cations.

## Conclusions

The second order Jahn-Teller effect determines the double-well local potential for $Cu^+$ cations and the three-well potential for $In^{3+}$ cations inside the structural layers of $CuInP_2S_6$ crystal, as demonstrated by *ab initio* MD simulation. The temperature dependence of the spatial distribution of $Cu^+$ and $In^{3+}$ cations described in the quantum anharmonic oscillators model demonstrates an important role of both $Cu^+$ disordering and $In^{3+}$ displacive dynamics in peculiar properties of ferrielectric states.

Positive hydrostatic or uniaxial normal to the structural layers stress promotes the throw of $Cu^+$ cations into the vdW gap. Negative biaxial stress along the structural layers rearranges the local three-well potential for $In^{3+}$ cations and promotes the widening of the temperature range for the FI2 ferrielectric state with small spontaneous polarization that continuously appears at cooling from the paraelectric phase. This continuously raised contribution into spontaneous polarization is related to the central part of the triple *P-E* loop and it mostly follows from the displacive dynamics of $In^{3+}$ cations.

The peculiarities of polarization switching in $CuInP_2S_6$ crystals at different temperatures or under compressive or stretching external pressures correlate with the observed ones under the influence of the internal chemical pressure that appears at partial substitution of Cu by Ag or S by Se. The multiple polarization states, that appear as single, double or triple *P-E* hysteresis loops, can be interpreted as the sum of contributions from $Cu^+$ ordering and $In^{3+}$ shifting under influence of the electric field.

# Figures

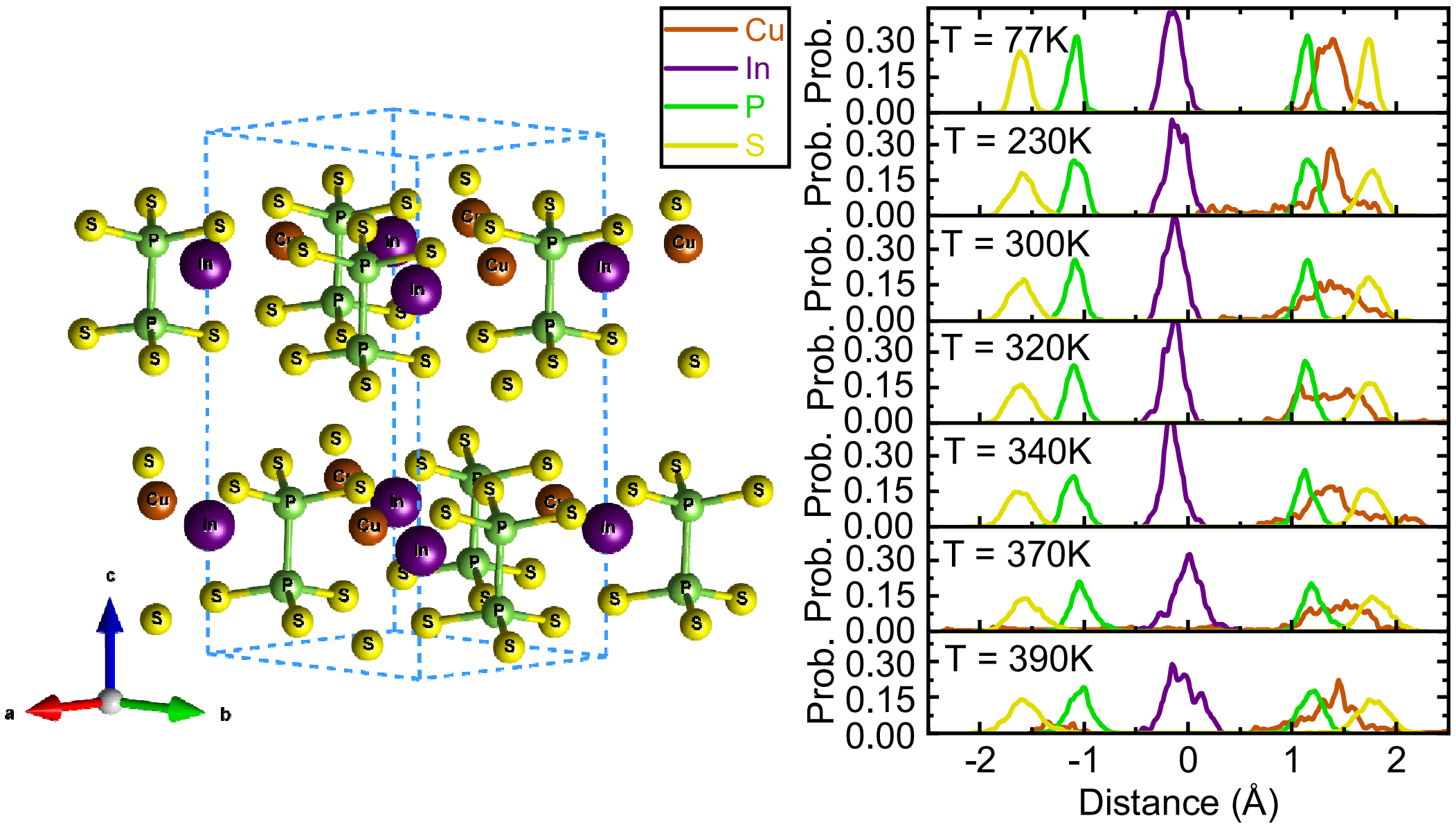

Figure 1. Modeling cluster of $CuInP_2S_6$ crystal with four primitive unit cells (left). Ion displacements from the plane of the structural layers averaged over the simulation time and for both layers for each temperature present in the figure (right). The probability is proportional to the time spent in the position and the depth of the potential at that point.

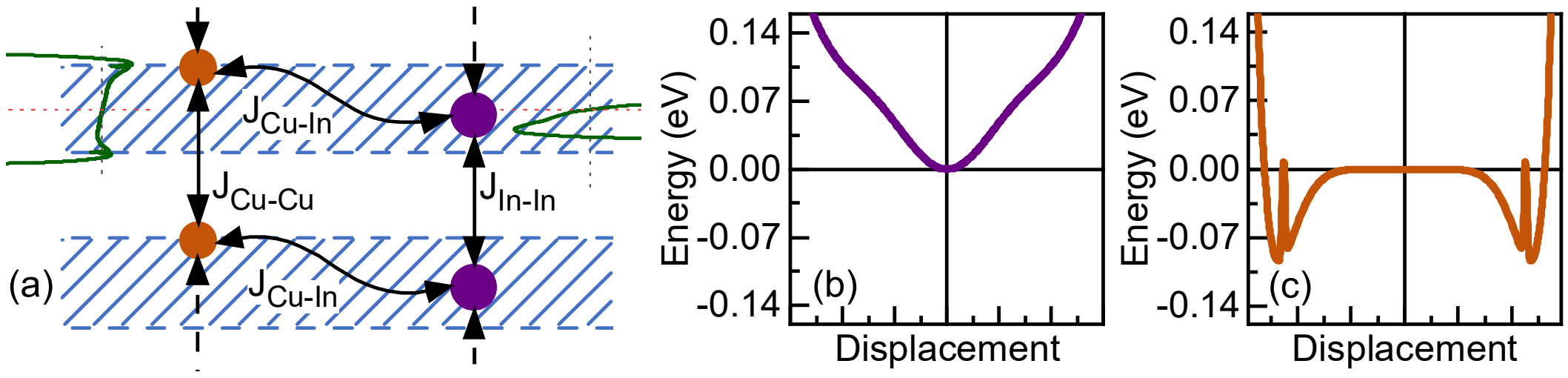

Figure 2. Quantum anharmonic oscillators model for the ferrielectric phase transition in $CuInP_2S_6$ crystal (a). The shape of the local potential energy landscape is shown for $In^{3+}$ cations (b) and for $Cu^{+}$ cations (c).

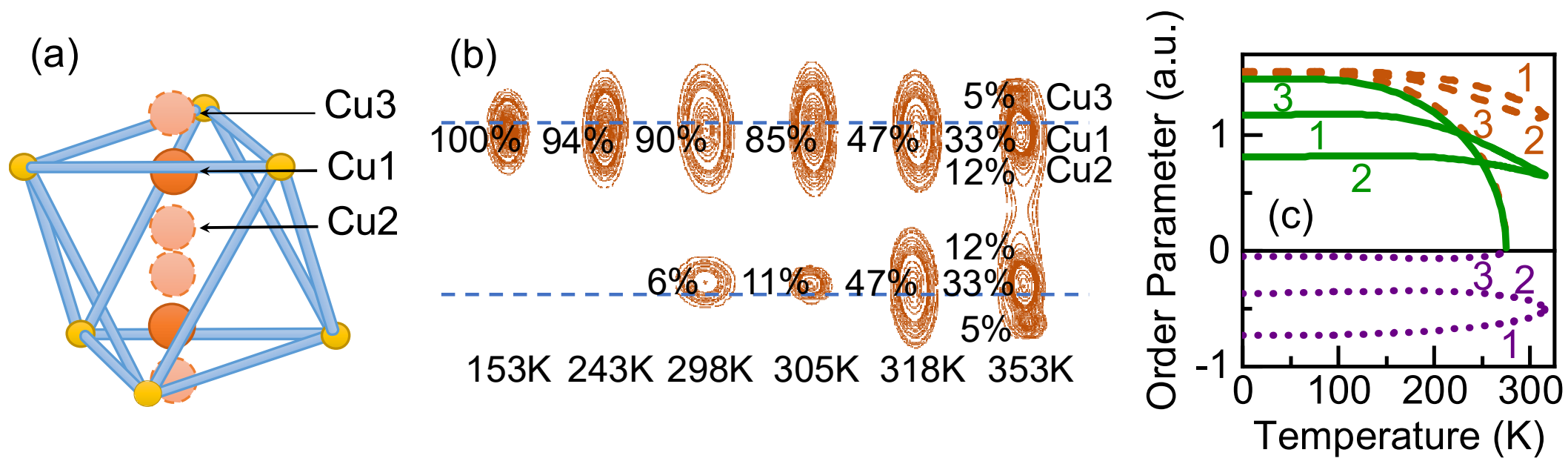

Figure 3. Sulphur octahedral cage with the various types of copper sites in the paraelectric phase: the off-center Cu1, the almost central Cu2, and Cu3 in the interlayer space (a), temperature dependence of the probability density contours for the spatial distribution of $Cu^+$ cations for $CuInP_2S_6$ crystal according to the electron diffraction data [1], the dashed blue lines indicate the upper and lower sulphur planes of a single layer (b), temperature dependence of the order parameter (spontaneous polarization) of $CuInP_2S_6$ crystal (green solid line) as a sum of oppositely oriented polarization contributed by $Cu^+$ cations (brown dashed line) and $In^{3+}$ cations (violet dotted line). The stable solution is related to the first order phase transition at $T_c$ = 315 K and marked with number **1**. Two metastable solutions are denoted by numbers **2** (related to the first order phase transition) and **3** (related to the second order phase transition at $T_0$ = 275 K) (c).

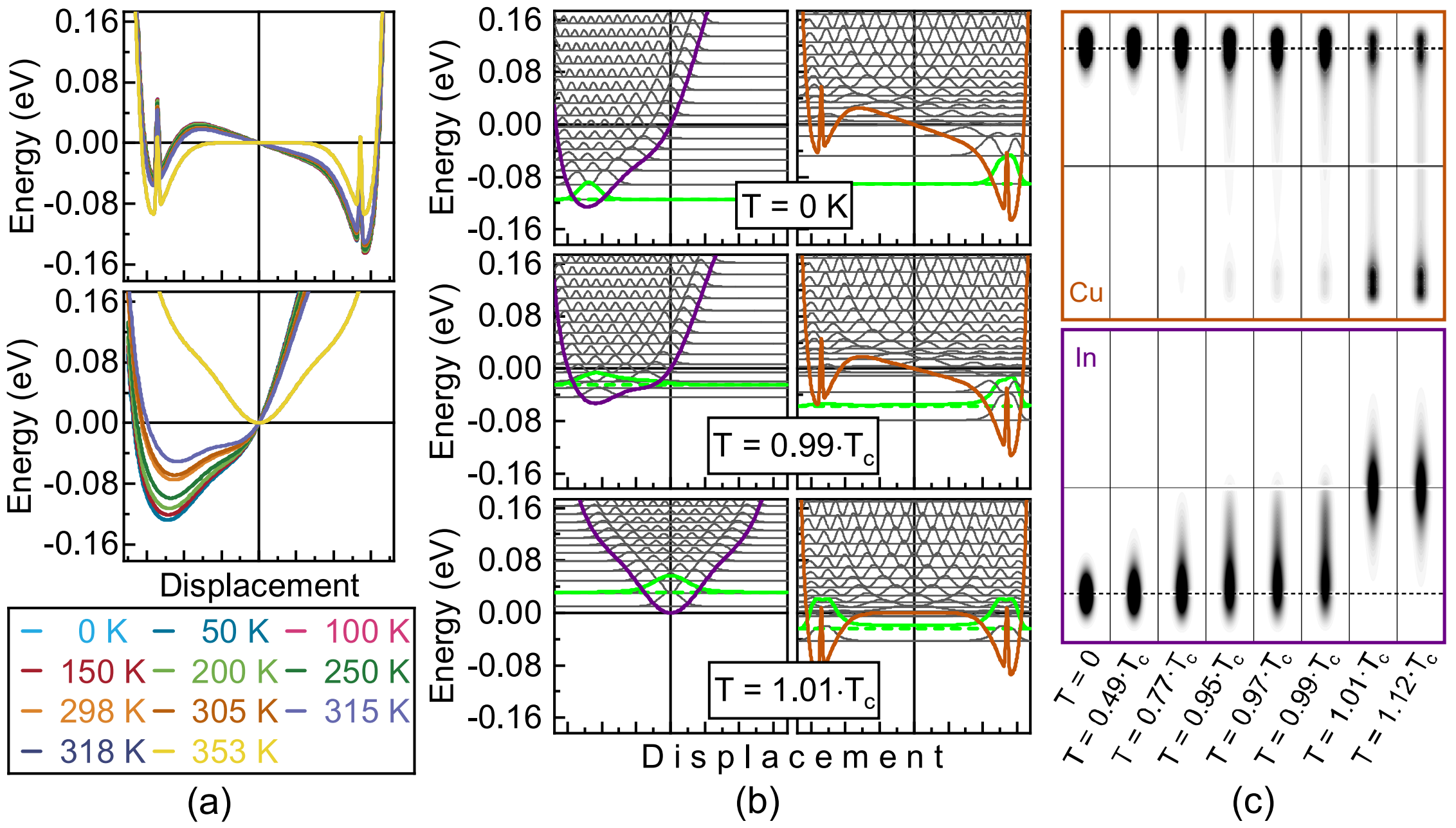

Figure 4. Temperature transformation of the double-well potential for $Cu^+$ cations and the three-well potential for $In^{3+}$ cations for stable solution **1** across the ferrielectric-to-paraelectric first order phase transition ($T_c$ = 315 K) (a), sets of eigen energies of levels

and their wave functions for the corresponding quantum anharmonic oscillators (b), probability density contours for the spatial distribution of copper and indium cations for solution **1** (c).

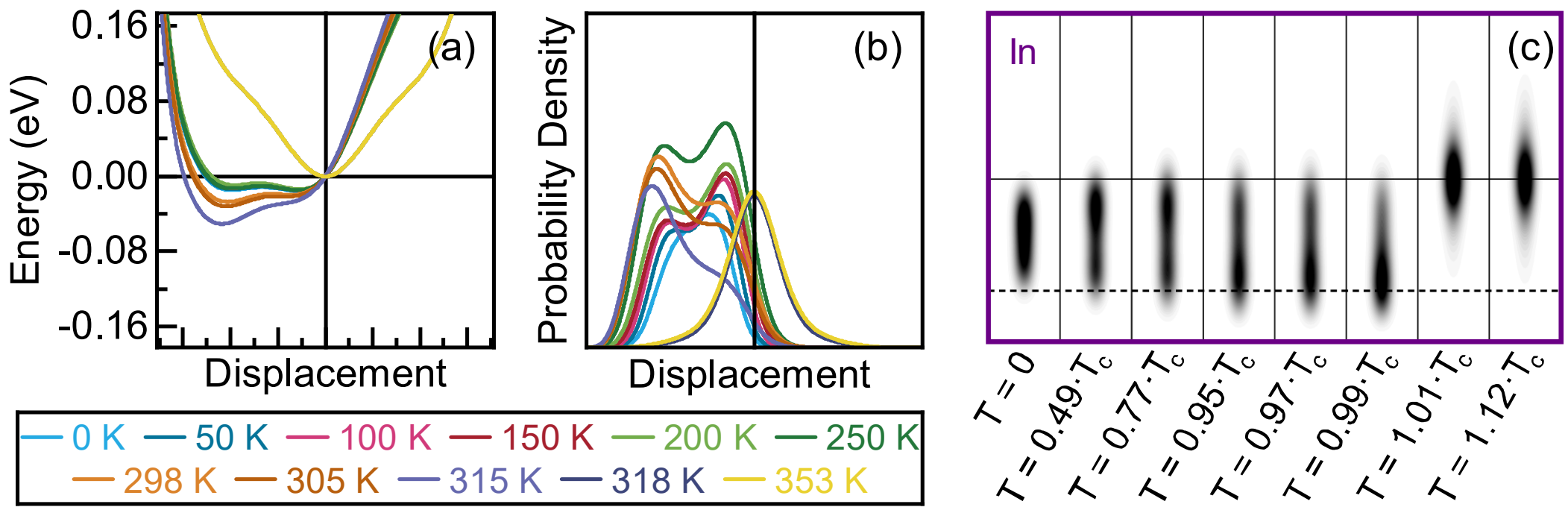

Figure 5. Temperature transformation of the three-well potential for $In^{3+}$ cations (a), probability density of their displacements relative to the center of the structural layers for metastable solution **2** across the ferrielectric-to-paraelectric first order phase transition ($T_c$= 315 K) (b), probability density contours for the spatial distribution of indium cations for solution **2** (c).

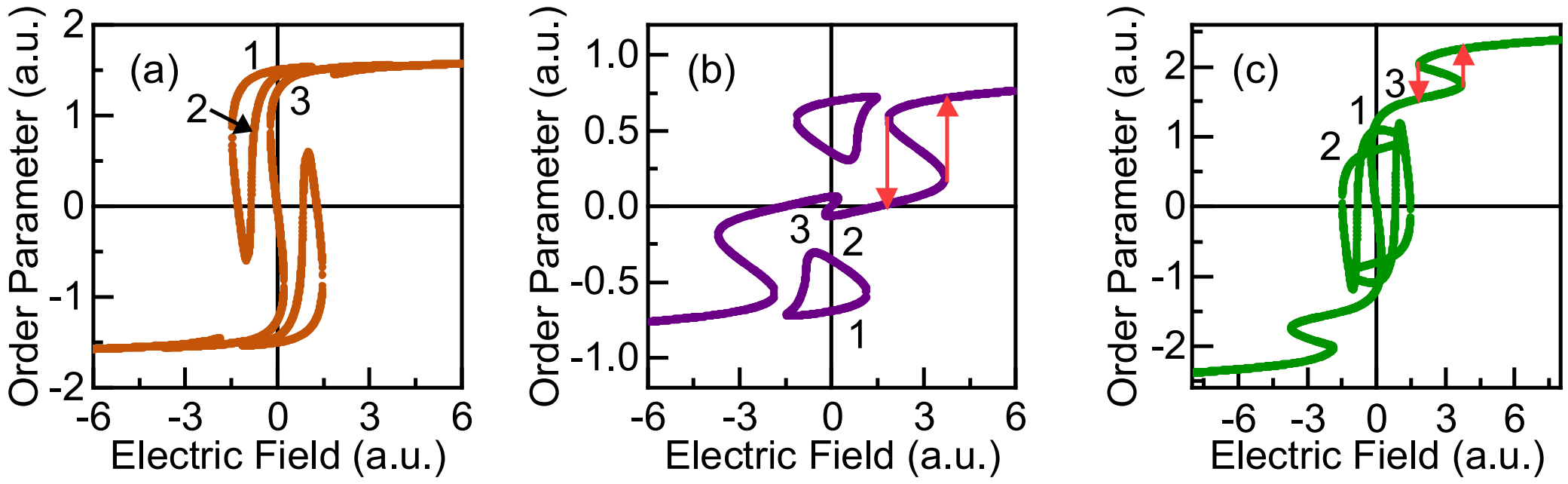

Figure 6. Contributions of stable **1**, metastable **2,** and **3** solutions donated by copper (a) and indium (b) sublattices to the total polarization hysteresis loop (c) in the ferrielectric phase of $CuInP_2S_6$ crystal at 200 K. Related to solution **3** displacement of indium cations approaches zero and takes the opposite orientation at a higher electric field - a transition from the ferrielectric to ferroelectric state occurs (shown by red arrows).

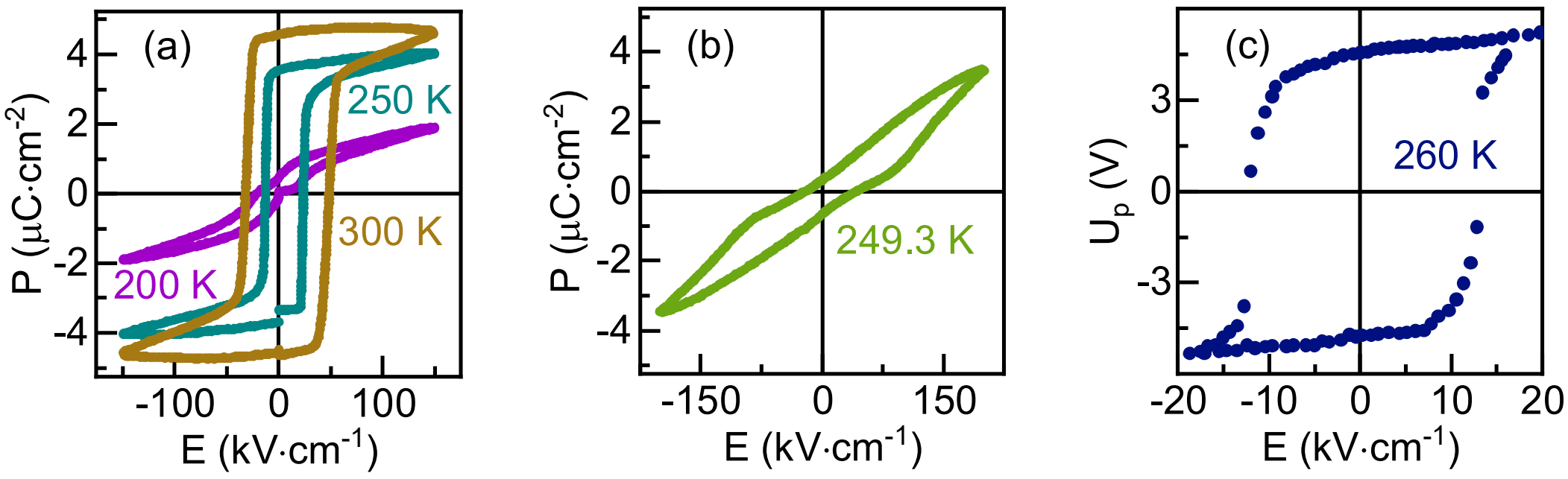

Figure 7. Polarization switching characteristics: *P-E* hysteresis loops for $CuInP_2S_6$ [17] (a) and $CuInP_2(Se_{0.02}S_{0.98})_6$ [16] (b) crystals; the DC field dependence of the piezoelectric signal for $Ag_{0.1}Cu_{0.9}InP_2S_6$ crystal (c) [18].